# Transition from 12 to near- 24 hours' glucose circadian rhythm on relaxation of a hyperglycemic condition.

Baruch Vainas

Soreq – NRC, Yavne 81800, Israel

## Abstract

A composite, exponential relaxation function, modulated by a periodic component, was used to fit to an experimental time series of blood glucose levels. The 11 parameters function that allows for the detection of a possible rhythm transition was fitted to the experimental time series using a genetic algorithm.
It has been found that the relaxation from a hyperglycemic condition following a change in the anti-diabetic treatment, can be characterized by a change from an initial 12 hours' ultradian rhythm to a near - 24 hours' circadian rhythm.

## Introduction

The present study describes, what appears to be a complex relaxation pattern of a hyperglycemic condition, upon change of treatment.
The relaxation appears to follow an exponential envelope, modulated by periodic ultradian and circadian periodic rhythms.
The apparent daily rhythmic pattern had been noticed from the very beginning of the blood glucose monitoring process, that had been initiated as the result of a unscheduled measurement of unusually high non-fasting glucose levels (maximal values approaching 300 mg/dL) detected, while the usual levels up to that point in time were about half that much, or less.
It had been noticed that minima in glucose levels seem to cluster at early mornings (~ 4-5 am) and afternoon (~ 4-5 pm), while maximum levels occur most frequently at around 10-11 am, and 9-11 pm, suggesting an apparent 12 hours rhythm (two maxima and two minima in 24 hours). As the relaxation to more moderate glucose levels had progressed, the rhythm had changed to near 24 hours circadian.
It was this observation that motivated an analytical study using a long time exponential relaxation function modulated by a periodic function, with its amplitude ("swing") relaxing in an exponential manner as well. The suggested analytical function was then fitted onto the experimental glucose level data using a genetic algorithm procedure. This has enabled the determination of important relaxation parameters, such as the relaxation time constant, and the glucose level expected to be reached at long times. The 12 and near 24 hours rhythms were then established from fitting the analytic function as described below.

Measurements were made using Ascensia ELITE XL blood glucose self test meter by a 59 years old DM type II, non insulin dependent diabetes (NIDD) patient.

Diabetes was diagnosed about 9 years before this study took place, and since then it was treated with a small dose of oral Glibetic tablet (Glibenclamide, 2.5 mg daily). Both the fasting glucose and A1C (HbA1C) values, of 116 mg/dL, and 6.0 respectively, last measured on July 2007, have been close, or within the norm, during all 9 years since the diagnosis of DM.

The patient is also being treated for high BP, with moderate doses of ACE, Ca, and Alpha blockers.

## Methods

Unusually high glucose level (275 mg/dL) was measured, by chance, on November 18, 2007 at 22:00, and regular recording of glucose levels started since then, 3-5 times in 24 hours. From Nov 19 through to Nov 22, the intake of Glibetic was increased to 5 mg daily, and a new antidiabetic tablet, Metformine 850 mg, was added. At first, Metformine was taken once a day, and after about two weeks, it was taken twice a day. This combination of oral antidiabetic medications was kept the same during the entire period of time analyzed here.

It is important to note that glucose levels were measured at unequally spaced time points. One day measurements could have been taken in 6:30, 11:45, and 22:05 time sequence, while the next day, there might have been 7:20, 12:30, 21:20, 2:00 (of the next day) time points. On detection of extreme values, a repeated sampling was made after about an hour. At any case, glucose readings were always taken, at least 2 hours after meals.

Data taken at unequal time intervals can't be used directly as an input to Fourier spectral analysis, if the determination of a possible periodic rhythm is postulated to be embedded into the long time relaxation process. A prior interpolation to get "missing" data should be needed, which could introduce estimation errors in the power spectrum.

Therefore, curve fitting algorithms were used, which do not require equally spaced time intervals.

### Curve fitting

A simple approach, studied here, consists of using a large collection of unequally spaced data, spread over many sequent time blocks (days, in the present case), rather than using interpolated values, and trying to use Fourier spectral density methods. This approach can be regarded to belong to the curve fitting variety.

A typical example of such curve fitting methods is the so called, "cosinor" [1] procedure for detecting circadian rhythms. For each pre-fixed period in the cosine function (starting from, say, 10 h, in 0.1 h increments, to about 25 h) it gives the weights of all the periodic components, and the goodness of fit [2] of the most significant one. Dominant rhythms can thus be found.

It is clearly a curve fitting method, using a simple trigonometric function to fit onto the data. The usual function form in this case needs just three parameters: the phase (called acrophase), the mean or level value (mesor), and the multiplicative constant, or amplitude, A.

The cosinor method does not include any non-periodic functionality, such as an exponential relaxation envelope, and neither can it automatically find different dominant rhythms in different regions along the time series.

These extra capabilities are possible with the genetic algorithm (GA) fitting method developed here. None of the fitting parameters (11 in total, including the time instant where rhythm transition can take place) are fixed in advance. GA is a global search algorithm that can find all these parameters, starting from random values.

While the relaxation process studied here can be seen as dominated by the initial, 12 hours rhythm, as shown below, the slower, overall rate dynamics of the exponential relaxation of the hyperglycemic condition on change of medications is most important. The long time relaxation dynamics is characterized by the time constant, and the long time limiting glucose level, given by the analytic fitting function. These two parameters give the important practical estimate as to how fast, and to what final glucose level could the medication initiated relaxation lead at the long run.

To address these two important parameters, the first part, $Y_1(t)$ of the overall relaxation function that will be tried to fit the data, is a pure exponential relaxation function given as:

$$Y_1(t) = a + b*exp(-t/c) \qquad \text{eq. 1}$$

Here, at long times, t, the simulated glucose level, $Y_1(t)$, will approach, a, which is the long time, relaxed value of the Y function. At, t = 0, the function, $Y_1(t) = a + b$, is the maximal, initial, value of the relaxation function.

The important time constant, c, can be used for the estimation of the so called "half life time" ($t_{1/2}$), which is the time expected for the Y to drop from its initial a + b value to a + (b/2), which is half way down, given by:

$$½ = exp(-t_{1/2} / c) \qquad \text{or,} \qquad t_{1/2} = c*\ln(2) = 0.693*c \qquad \text{eq. 2}$$

In the curve fitting result that leads to the 11.9 h and 24.2 h rhythms (see details below), the mean glucose level does fall halfway down at about ~0.7*544 = 380.8 hours, or, in 380.8/24 = 15.9 days, which could be considered as a moderately fast response to the change in treatment. The expected, long time limiting glucose level, tends towards 98 mg/dL, as indicated by the value of a, meaning that of normal mean glucose levels can be expected on the long run.

The periodic, modulation of the long time relaxation process is given by a second part, $Y_2(t)$, in the overall relaxation function. This part is combined with the main exponential relaxation $Y_1(t)$, and it has got its own exponential relaxation term, that determines the rate at which the amplitude of the periodic function relaxes with time.

The rhythmic part is given by:

$$Y_2(t) = (k + l*exp(-t/m))*cos(2*Pi((t/n) + p)) \qquad \text{eq. 3}$$

Note the first exponential relaxation expression in the first parentheses, and the period, n, phase, p.
The complete relaxation function is just,

$$Y(t) = Y_1(t) - Y_2(t) \qquad \text{eq. 4}$$

In this form, the pure exponential relaxation is modulated by the periodic part by simply subtracting it from the long time envelope of $Y_1(t)$. The GA then finds the best set of parameters defining both $Y_1(t)$ and $Y_2(t)$, so that their combination according to eq. 4, provides the best fitting to the experimental time series.
After glucose monitoring progressed to longer times, it has been realized that the initial 12 hours rhythm in $Y_2$ could be changing to longer period rhythms. To allow curve fitting for this possible rhythm transition, three additional parameters have been added to the 8 parameters being used so far. These three additional parameters are the following:

tr – the point on the time axis, where rhythm changes, from its preceding one (that had been characterized by parameters, n, and, p, in eq. 3).
mu – the parameter replacing, n, after passing time tr.
ph – the parameter replacing, p, after passing time tr.

There are now 11 adjustable parameters characterizing the two rhythms fitting function, Y(t). That was found as too many for a commercial curve fitting package – TBCurve (for DOS) which could handle only 8 parameters for user defined fitting functions.

A simplified, 8 parameters fitting function (with a single rhythm) was then tried with the TBCurve software on the initial (in time) part of experimental data. A clear identification of the 12 hours period was found.
Similarly, the dominant, 12 hours rhythm was clearly identified ($p < 0.0001$) using the cosinor method [2]. While cosinor procedure does show the longer period rhythms in the entire time series (that include 19.4 and 24.2 hours contributions), the transition point to a longer time rhythm at the last part of the time series can not be identified, and the overall, long time exponential relaxation can not be determined with cosinor, either.
The genetic algorithm has been chosen to allow for a large number of parameters to be fitted to an analytic function simulating the experimental data, including relaxation and the location in time of the rhythm change, while avoiding local minima.

**Genetic Algorithm and the cost function**

One of the search methods that can look for a global minimum or maximum, rather than falling into local extreemum points, is the genetic algorithm search that does not use seed values. Rather, it uses random values from the very beginning, and keeps "refreshing" the trial parameters population by what is known as the mutation operator, and forced randomization of relatively "fit" but "old" parameters sets (called "chromosomes" in genetic algorithms speak).
This way, GA could find a global minimum in terms of the 11 parameters (in the present case) of the "cost" function. The cost function is defined as the mean "distance", or difference, between experimental values of glucose at a given time points, and the corresponding values given by the fitting, analytic relaxation function, evaluated at the same time instants as the real data.
The GA procedure used in this study can be described in a simplified way as:

1. Initiation – create a population of a large number of 11 element groups (or, vectors/chromosomes). Each element is populated by a random number.
2. Evaluation – evaluate the cost function (distance from experimental points) for each group of elements, which are the candidate parameters of the fitting function. The cost function in this case is the root mean square (RMS) value of the differences between the experimental and fitted points along the entire time series, as given by the fitting function, $Y(t)$.
3. Update the best chromosome (the leader) and protect it from the action of genetic operators that follow in the next step. Also update the parameters of all chromosomes that have been improved, meaning – decreased their cost function value relative to their state before the last genetic operation took place.
   Improved vectors that are not the leader (improved, but not reached the low cost function of the leader) will be protected from change, but only for a finite number of cycles. Any protected vector can donate the values of its elements in a possible crossover operation, but it keeps its original values (not "corrupted by inferiors") while still being protected.
4. Do either a point to point crossover operation between randomly chosen pairs of chromosomes (exchanging one value, or more, of vector elements having the same index in vectors), or perform a complete randomization (mutation) of all vectors except the leader. Protected vectors (except of the leader) that haven't improved any further, after a predetermined number of steps get their protection lifted, and can be changed by these genetic operations. This is what is known as "reheating".
   Vectors that did not improve, or got worse, are not protected against changing their elements. They are exposed to modification with a probability related to their decrease in fitness relative to the previous state. Like in simulated annealing (SA) algorithm, this probability is given by a Boltzmann exponential factor. The present method can therefore be classified as a genetic-annealing type of global search algorithms.
5. Goto step No 2, and end cycles when reaching the predetermined number of cycles.

The output of the GA is the elements of the leading vector. It is a group of 11 parameters that gives the best fitting analytic function, $Y(t)$, that have its values at times, t, when the experimental data been measured, as close to the experimental glucose levels as the algorithm could get.

## Results and Discussion

The raw data is given in Fig. 1 below. The short vertical lines at the lower edge of the figure denote 24 hours intervals on the time axis.

Fig. 1

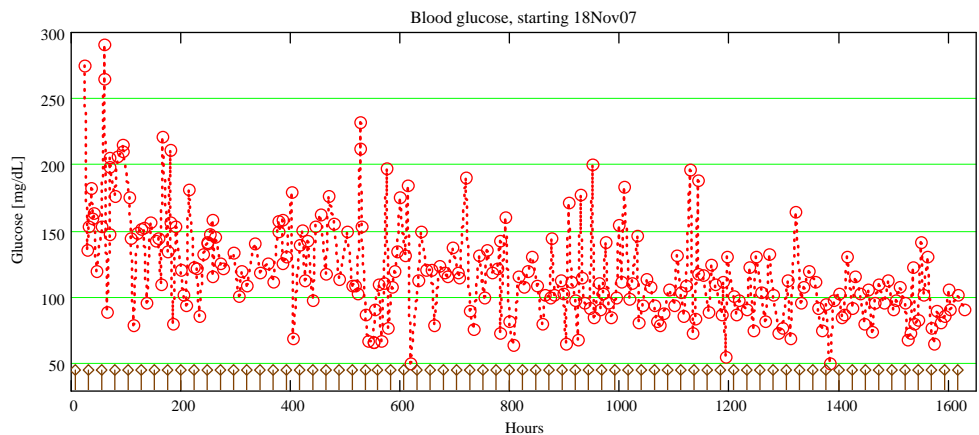

Two dominant curve fitting solutions have been obtained, based on the data available thus far. The first is – initial rhythm of 12 h, changing to 19.4 h, and the second – initial rhythm of 11.9 h changing to 24.2 h.
The most recent set of fitting parameters, derived from the GA for 248 data points (spread over total time of 1626 hours) is as follows:

For the 11.9 h / 24.2 h curve fitting result:
a = 98, b = 64, c = 544 (the long time exponential relaxation envelope)
k = 8, l = 69, m = 192, n = 11.9, p = 0.59 (exponentially relaxing periodic function)
tr = 305.1, ph = 0.94, mu = 24.2 (rhythm transition location, new phase and period)

For the 12 h / 19.4 h curve fitting result:
a = 102, b = 70, c = 356 (the long time exponential relaxation envelope)
k = 13, l = 71, m = 112, n = 12, p = 0.65 (exponentially relaxing periodic function)
tr = 800.1, ph = 0.58, mu = 19.4 (rhythm transition location, new phase and period)

Note that the units of the parameters are contextual to their meaning in the equations above. Thus, a, b, k, l are glucose concentration in mg/dL, c, m, n, mu, tr are in hours, while p, ph are pure numbers denoting the phases as fractions of 2Pi radians.
These particular sets of parameters define an analytical function, Y(t), that when compared to real values of blood glucose at the experimental time points, produce a root mean square error (RMS) of  ~ 30 mg/dL. The analytical function is designed to deal with the dominant trends. The function is derived by minimizing the RMS – the cost function, between data and function. The rather low value of RMS (in comparison with several large, ~50-100 mg/dL swings found in data) suggests that a large portion of data points and the corresponding points on the function are in close proximity.

That can be seen by inspecting Fig. 2 below (split into two non-overlapping parts, a and b, for clarity) which shows the individual experimental data points (circles on thin vertical lines) superimposed on the exponentially relaxing periodical 11.9 h / 24.2 h fitting function.

Fig 2a

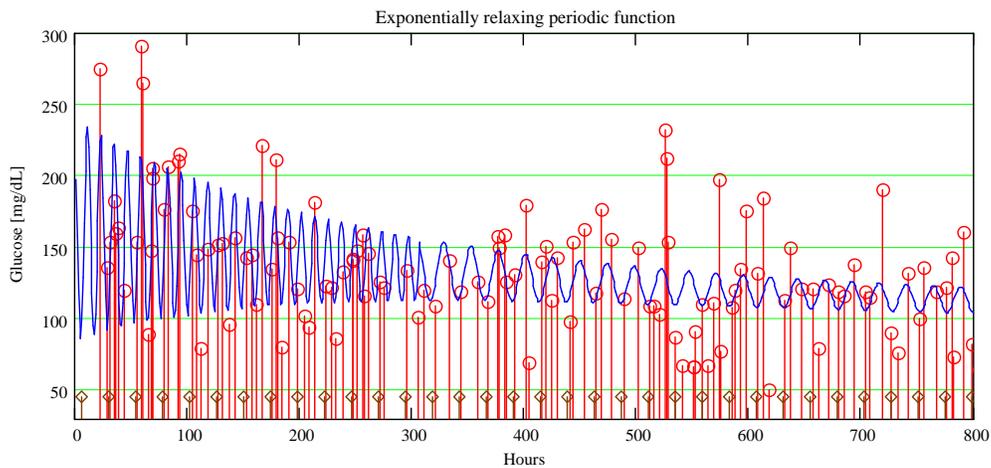

The continuation of this time series:

Fig 2b

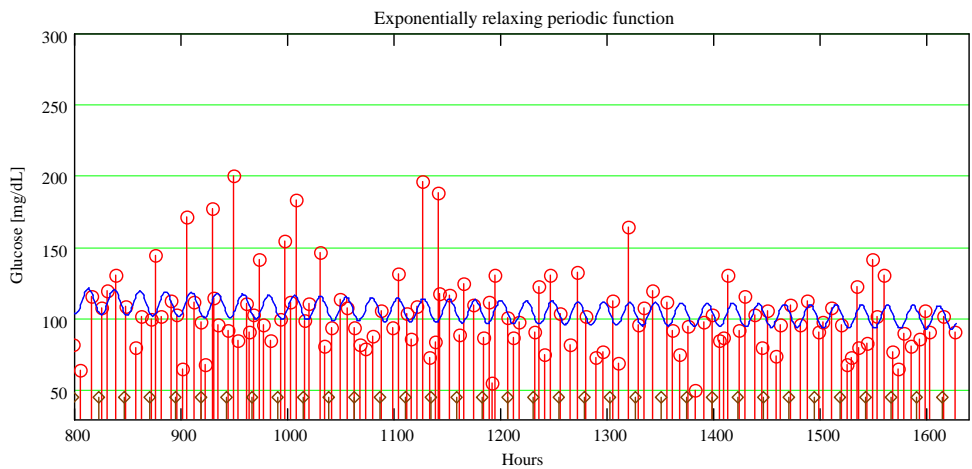

The corresponding figures for the 12 h / 19.4 h curve fitting are in Fig 3 below.

Fig 3a

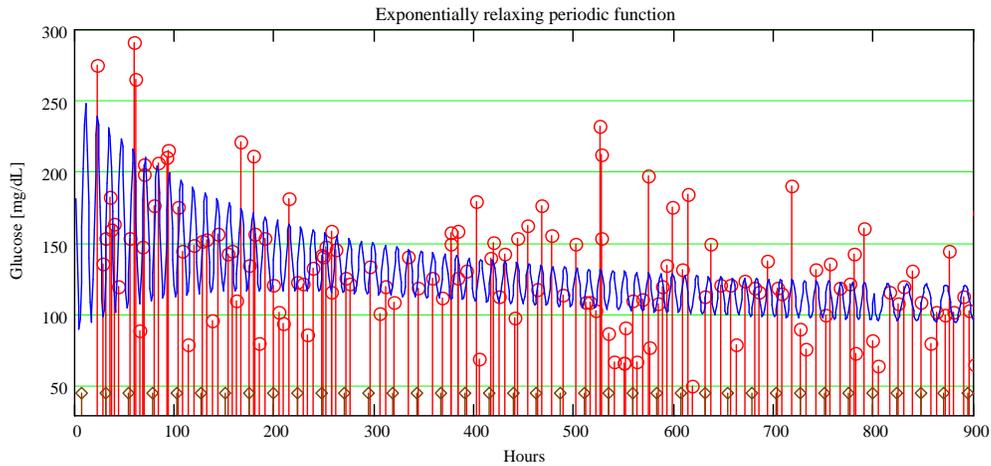

The continuation of this time series:

Fig 3b

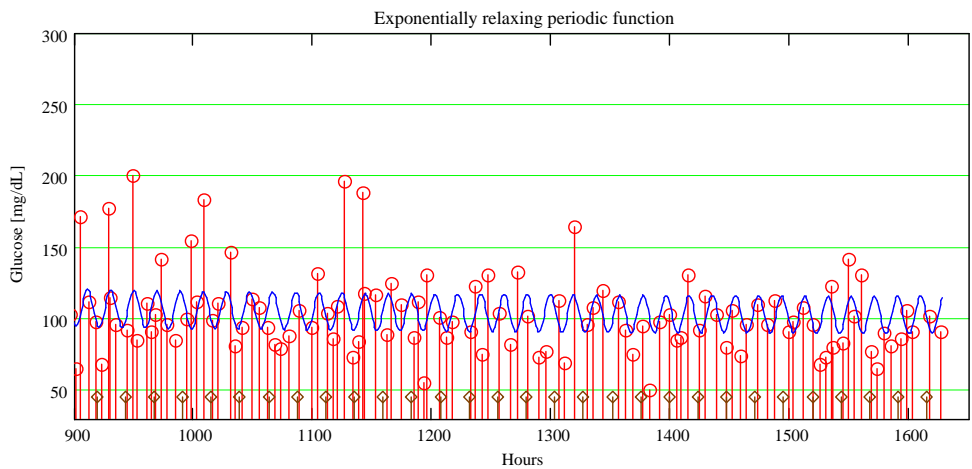

Note the different time axis arrangement of the two parts of Fig 3a and 3b to clearly show the rhythm transition around 800 hours.
Of a particular interest, one can notice the close correspondence between the location in time of the extreme points (of hypo, and hyperglycemic values) in data, and the function fitted. Thus, most hypo values (see for example the 291 mg/dL at 59 h data point) coincide in time with a sinus maximum of the periodic fitting function, Y(t). The correspondence is most evident in the initial 11.9 hours rhythm region, up to the point of 305 hours on the time axis (for the 11.9 h / 24.2 h fitting, in Fig. 2) , where the transition to the longer, circadian-like rhythm, takes place. Similarly, the clear correspondence can be noticed in the initial 12 h rhythm region, up to the 800 h point in time, for the 12 h / 19.4 h curve fitting, shown in Fig. 3.

The most significant parameters are found in the following categories:
 1. The long time limit of the relaxation envelope, a = 98 - 102 mg/dL, which can suggest that the medical treatment response can be expected to lead to an acceptable result.
2.  Time constant of the long time relaxation envelope is, 544 hours, for the 11.9 h / 24.2 h fitting function. The half life time is then, ~0.7*544 = 380.8 hours or, 380.8/24 = 15.9 days. It means that the initial high mean levels of glucose dropped to half their value in 15.9 days, which shows that the treatment became effective in a reasonable time.
3. The location (parameter, tr) on the time axis of the change in rhythm from 12 hours (parameter, n) to ~20, or ~24 hours. This has both practical and theoretical value. Thus, the treatment at the initial stage might benefit from a tighter and more frequent scheduling of medications intake, based on the location in time of the extreme values of glucose levels.

The rhythm's transition itself can be discussed in the context of non-linear dynamic systems where periodic trajectories can undergo what is known as period doubling under imposed change of control parameters, and even leading to chaotic characteristics under extreme values of such parameters. The relaxation to the more common circadian rhythm upon the relaxation of the initial state of hyperglycemia, can suggest the transition from period doubling back to the normal state, or close to it.

One of main criticisms of the present study can be the fact that the raw data used is not compensated for times of meals, physical effort, rest, and some other important factors.

While glucose was always sampled 2-4 hours after meals, or in the morning, after fasting, the ideal case would be a constant food intake, for example. However, the fact that measured glucose values were almost always higher at about 7-8 AM (just before breakfast), than those measured about two-three hours before that, can suggest that given a regular pattern of meals, the glucose levels at some time points are less dependent on meals than other time points. The same can be noted on the relatively high glucose levels at around 10-11 PM, which is quite distant from the dinner, usually taken at around 6-7 PM.

It is important to note that glucose measurements were taken at a large spread of times. The very high statistical significance of the initial, 12 hours rhythm, $p < 0.0001$, can indicate that this result is indeed significant, given the wide spread of measurement times.

The circadian 24 h rhythm is known to exist in non-diabetics [3]. One of its expressions is the temporary rise in glucose levels in the evening. The term "afternoon diabetes" was coined to describe this phenomenon because of the increased potential of a false-positive diagnosis of diabetes in the afternoon, as compared with the morning. This "afternoon diabetes" was shown to be a true effect of the time in the day, independent of fasting details [3]. This phenomenon of the true effect of the time is similar to the consistently higher levels of glucose measured at around 8 AM, relative to those at 5 AM, and high levels at 10 PM, as observed with the 12 h rhythms described above.

 The important question as to why the best fit in the present study indicates an approach to a circadian rhythm, with a 19.4 h or 24.2 periods (the two dominant curve fitting solutions) rather than to expected 24 h period is a subject for a further study. At the present time, it can be noted that apparently the non-harmonic, 19.4 h rhythm (unlike the 12 h rhythm) can suggest a possible non-stationary effect of a non completed process of relaxation to the circadian, 24 h rhythm.